\documentclass[aps,onecolumn,prb,floatfix,longbibliography, superscriptaddress,showpacs]{revtex4-2}
\usepackage[utf8]{inputenc} 

\usepackage[english]{babel}
\AtBeginDocument{}

\usepackage{float}
\usepackage{array}
\usepackage{ragged2e}
\newcolumntype{P}[1]{>{\RaggedRight\arraybackslash}p{#1}}

\usepackage{graphicx}
\usepackage{hyperref}
\usepackage[dvipsnames]{xcolor}
\usepackage{bm}

\usepackage{physics}
\usepackage{braket}
\usepackage{amsfonts}
\usepackage{amsmath}
\usepackage{amssymb}

\hypersetup{
	breaklinks = {true},
	citecolor = {blue},
	colorlinks = {true},
	linkcolor = {red},
  pdfstartview={FitH},
  linkcolor=NavyBlue, 
	citecolor=Maroon,   
	filecolor=NavyBlue, 
	urlcolor=NavyBlue   
}

\usepackage[normalem]{ulem}

\begin{document}

\title{Single-Q and Double-Q magnetic orders: A Theoretical Analysis of Inelastic Neutron Scattering in a Centrosymmetric Structure}
\author{Artem O. Nosenko}
\affiliation{Leibniz Institute for Solid State and
	Materials Research IFW Dresden, Helmholtzstra\ss e 20, 01069 Dresden, Germany}
\author{Dmitri V. Efremov}
\affiliation{Leibniz Institute for Solid State and
	Materials Research IFW Dresden, Helmholtzstra\ss e 20, 01069 Dresden, Germany}
\date{\today}

\begin{abstract}
	Recent discoveries of multi-\textbf{Q} magnetic structures in centrosymmetric
	compounds have stimulated growing interest in their microscopic origin and observable properties.
	Here, we calculate the dynamical magnetic structure factor for a double-\textbf{Q} magnetic
	structure and compare it with that of a single-\textbf{Q} configuration.
	  Our results demonstrate how inelastic neutron scattering can distinguish the double-\textbf{Q} state from competing single-\textbf{Q} states in this square-lattice setting.
\end{abstract}
\maketitle
\section{Introduction}
In recent years, significant effort has been devoted to the study of
non-collinear magnetic structures \cite{Tokura2010, Kimura2012,
	Nagaosa2013, Jiang2017}. Among these, topological spin textures such
as skyrmions have attracted particular attention due to their
potential applications in next-generation computing technologies
\cite{Tokura2021, Mishra2025}. More recently, interest has grown
following the demonstration that complex spin configurations, such as
skyrmion lattices \cite{Hamamoto2015, Goebel2017, Wang2020,
	Kurumaji2025, Back2020}, can form spin-Moiré superlattices
\cite{Shimizu2021}, leading to uniquely modulated electronic
behavior. This discovery has opened up a new field of research, with
the exploration of novel applications of topological spin structures
emerging as a key focus.

The majority of the investigated systems that exhibit
multi-\textbf{Q} magnetic structures are non-centrosymmetric. In
these systems, the Dzyaloshinskii-Moriya interaction (DMI)
\cite{Dzyaloshinsky,Moriya} is a key factor in the formation of
various noncollinear spin structures. An external magnetic field can
change a single-\textbf{Q} magnetic structure, turning it into a
multi-\textbf{Q} one \cite{Bogdanov2006,Muehlbauer2009}.

In centrosymmetric materials, frustrated interactions can result in
various noncollinear structures
\cite{Vllain1959,Kaplan1959,Yoshimori1959}. In particular, several
theoretical studies have predicted the formation of multi-\textbf{Q}
magnetic structures in centrosymmetric systems under specific
conditions \cite{Okubo2012, Hayami2017, Batista2016,Leonov2015}.
Despite these theoretical advances, clear experimental evidence for
the existence of topological multi-\textbf{Q} structures in
centrosymmetric compounds has only emerged very recently.

In the past few years, a limited number of centrosymmetric materials
have been reported to host skyrmion-lattice phases. These include
Gd$_2$PdSi$_3$
\cite{Kurumaji2019a,Zhang2020,Spachmann2021,Paddison2022,Gomilsek2025},
 GdRu$_2$Si$_2$ \cite{Khanh2022,Wood2023,Paddison2024}, EuAl$_4$
\cite{Takagi2022,Gen2023,Korshunov2024}, EuGa$_2$Al$_2$
\cite{Moya2022,Vibhakar2023}, Sr$_3$Fe$_2$O$_7$
\cite{Kim2014,Andriushin2024} and SrFeO$_3$
\cite{InosovIshiwata2020,Andriushin2025}. It was shown that despite
the presence of inversion symmetry, they can host multi-$\mathbf{Q}$
magnetic ordering and topological phases such as skyrmion and
hedgehog lattices. Experimental studies have revealed the most
important signatures of the multi-\textbf{Q} magnetic structures such
as the topological Hall effect and magnetoresistance anomalies
associated with spin fluctuations. 

Recently, the magnetic state of Sr$_3$Fe$_2$O$_7$ has been examined
in inelastic neutron scattering experiments \cite{Kim2014,
	Andriushin2024}. Sr$_3$Fe$_2$O$_7$ crystallizes in a
layered Ruddlesden–Popper structure composed of double perovskite
layers of corner-sharing FeO$_6$ octahedra. Upon cooling, it
undergoes a continuous transition from a high-temperature metallic
phase to a low-temperature insulating state at $T_{MIT} = 340 $K,
followed by the onset of noncollinear magnetic order below $T_N =115$~K 
\cite{Peets2013, Kim2014}. The system exhibits pronounced
two-dimensionality and strong magnetic frustration. Recent systematic
investigations have revealed a complex magnetic phase diagram that
may host both single-\textbf{Q} and multi-\textbf{Q} magnetic phases,
based on the magnetic structure factor \cite{Andriushin2024}. This
observation motivates a detailed theoretical analysis of their
neutron scattering signatures.
However,
directly confirming 
multi-\textbf{Q} magnetic
structures is still lacking.

The distinction between multi‑$\textbf{Q}$ and single‑$\textbf{Q}$ magnetic orders by means of their spin‑wave spectra is a well‑known concept.  Early spin‑wave calculations for a triple‑$Q$ structure showed clearly different excitation branches compared with the corresponding single‑$\textbf{Q}$ case  \cite{Jensen1981}.  More recent work on powder inelastic‑neutron‑scattering data has demonstrated the practical usefulness of this approach for identifying multi‑$\textbf{Q}$ order, and a comprehensive theoretical analysis of triple‑$\textbf{Q}$ dynamics on a triangular lattice has further clarified the generic spectral fingerprints 
         \cite{Paddison2024_2,Motome2021, Park2025}.

In the present study we focus on a square‑lattice model with anisotropic exchange interactions that is directly relevant to the layered compounds Sr$_2$FeO$_4$ and Sr$_3$Fe$_2$O$_7$.  By constructing the complete phase diagram in the anisotropy parameters plane and performing linear‑spin‑wave calculations of the inelastic neutron cross-section for the competing single‑$\textbf Q$ (helix, spiral) and double‑$\textbf Q$ configurations. We identify a set of robust, experimentally accessible signatures of double‑$\textbf Q$ order—most prominently a roton‑like magnon mode, the suppression of the outer vertical branch, and a gapless phason at both propagation vectors.  These features persist after realistic domain averaging and magnon damping are taken into account, providing a concrete guide for future inelastic neutron‑scattering experiments on the iron‑oxide materials of interest.

The paper is organized as follows. After a brief introduction, we
construct the magnetic phase diagram as a function of the anisotropy
parameters. We then present a detailed analysis of the dynamical
structure factor for the magnetic phases identified in the phase
diagram. Finally, we conclude with a discussion and summary of our
key findings.

\begin{figure*}[t!]
	\begin{minipage}[t]{0.49\textwidth}
		\centering
		\includegraphics[width=\columnwidth]{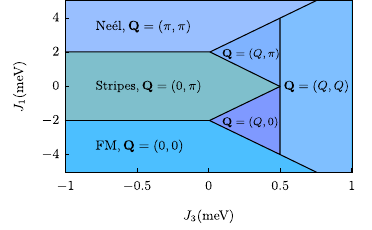}
		\caption{Ground state of the isotropic $J_1\!-\!J_2\!-\!J_3$
			Heisenberg model of a classical spin system.
			The states are described as helices with the corresponding
			propagation vectors $\mathbf{Q}$. Parameters: $J_2=1$ meV. }
		\label{fig:LTPhaseDiagram}
	\end{minipage}
	\hfill
	\begin{minipage}[t]{0.49\textwidth}
		\centering
		\includegraphics[width=\columnwidth]{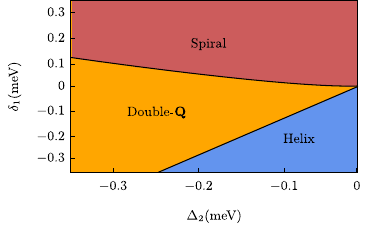}
		\caption{Ground state of the anisotropic Heisenberg model. The diagram reveals several magnetic phases:
			Spiral --- a spiral phase with spin rotation in XY plane,
			Helix --- a "proper screw" rotation perpendicular to $(1,1)$
			and Double-\textbf{Q} configurations. The exchange parameters are $J_1 = -7.2$~meV, $J_2 = 1.05$~meV, and $J_3 = 2.1$~meV.}
		\label{fig:PhaseDiag}
	\end{minipage}
\end{figure*}

\section{The model}
We investigate the following Hamiltonian defined on a square lattice:
\begin{equation}
	\begin{split}
		H  = & \sum_{\scriptscriptstyle  ij } \sum_{\scriptscriptstyle \alpha \beta } J_{ij}^{\alpha \beta} S_i^{\alpha}  S_j^{\beta}   , \\
	\end{split}
	\label{eq: Ham}
\end{equation}
where $i,j$  and $\alpha,\beta= x,y,z$ are  the site and   spin indices, correspondingly.
The interaction is parameterized by sets of three parameters $(J_\ell,\Delta_\ell,\delta_\ell)$ as  was proposed  for Sr$_3$Fe$_2$O$_7$ in \cite{Andriushin2024}:
\begin{equation*} \label{ExchangeXY1}
	\begin{split}
		J_{ij}^{\alpha\beta} & = J_\ell \delta^{\alpha\beta} + \Delta_\ell \left(\hat{r}_{ij}^{\alpha} \hat{r}_{ij}^\beta - \frac{1}{2}\delta^{\alpha\beta}\right) \mbox{~~for~~} \alpha,\beta = x,y, \\
		J_{ij}^{zz}          & = J_\ell + \delta_\ell,
	\end{split}
\end{equation*}
where  $\ell=1,..,3$ corresponds to the nearest-neighbors, next-nearest-neighbors, and third-nearest-neighbors interactions.
Here $\hat{r}_{ij} = (\mathbf{r}_i - \mathbf{r}_j)/|\mathbf{r}_i - \mathbf{r}_j|$, where $\mathbf{r}_i$ is the coordinate of the $i$-th spin. For $\Delta_{\ell}=\delta_{\ell}=0$ the model becomes the isotropic Heisenberg model.

\begin{figure*}[!t]
	\centering
	\includegraphics{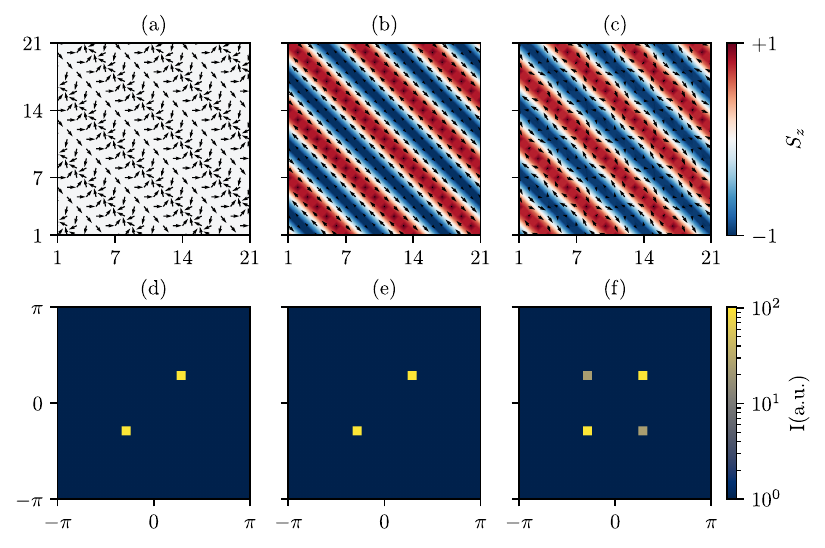}
	\caption{
		Spin structures (a-c).
		The in-plane components $S^x$ and $S^y$ are depicted as arrows and the out-of-plane component $S^z$ is presented by color:
		(a) Spiral,
		(b) Helix,
		(c) Double-\textbf{Q},
		(d-f) corresponding magnetic structure factors.
		Parameters for the Double-\textbf{Q} state:
		\( J_1 = -7.2\) meV, \( J_2 = 1.05\) meV, \( J_3 = 2.1 \) meV,
		\( \Delta_2 = -0.36 \) meV, and \( \delta_1 = 0.072 \) meV.
		Parameters for the spiral and helix states are given by
		$J_1 = -2J_2\cos\theta - 4J_3\cos\theta$, with $\theta = 2\pi/7$ , $J_2 = 1.05$ meV, $J_3 = 2.1$~meV, $\Delta_{1,2,3} = 0$, and $\delta_{1,2,3} = 0$.}\label{fig:Spintexture.SSF}
\end{figure*}

\section{Magnetic phases}
First, we discuss magnetic structures within the isotropic Heisenberg
model Hamiltonian ($\Delta_{\ell} = \delta_{\ell} =0$, $\ell
	=1,..,3$). Spin configurations that minimize the classical energy can
be determined analytically, e.g. using the Luttinger--Tisza method. In
this regime, the system stabilizes in a single-\textbf{Q} magnetic
structure (helices), which is characterized by a propagation vector $
	\textbf{Q} = (Q, 0) $ (helix along $x$- axis or $y$-axis) and
$\textbf{Q} = (Q, Q)$ (helix along the diagonal)~
\footnote{The point
	$J_1=J_3=0$ is a singular limit of the model. In this case, the
	system decomposes into two independent sublattices coupled only
	through $J_2$, and the classical ground state is correspondingly
	degenerate, as the referee correctly notes. The stripe region in the
	phase diagram should therefore not be interpreted as representing a
	uniquely defined ordered state at this isolated point.}
. Here $
	4(J_2/2 + J_3) \cos(Q) + J_1 =0$ and we assume here and below the
lattice constant $a=1$. Because the Hamiltonian does not incorporate
spin–orbit interaction, it remains invariant under rotations of the
spin plane.

The results of the calculations are summarised in Fig.
\ref{fig:LTPhaseDiagram}. As can be seen from the phase diagram,
non-collinear structures are realized in the presence of frustrated
interactions, specifically when the nearest-neighbor interaction is
ferromagnetic ($J_1<0$) and the next-nearest-neighbor and
third-nearest-neighbor interactions are antiferromagnetic
($J_{2,3}>0$).

However, to explore more complex states with multiple $\mathbf{Q}$
vectors in the parameter space of $\delta_1$ and $\Delta_2$, where
the Luttinger--Tisza method is not applicable, the classical ground
state must be determined using numerical approaches such as Monte
Carlo simulations~\cite{JankeMinimization, NumericalOptimization}. We employ the Metropolis Monte Carlo algorithm on
spin lattices typically consisting of $21\times21$
sites~\cite{suppl}. Starting from a random spin
configuration at high temperature, the system is gradually cooled to
a low-temperature state, allowing it to relax into a candidate
low-energy configuration. To ensure that the resulting spin
configurations, in particular for multi-$\mathbf{Q}$ states, are not
artifacts of finite system size or boundary conditions, we perform
additional simulations for varying lattice sizes (see Fig.~S1 in the Supplemental Material~\cite{suppl}).

The spin configurations obtained from the Monte Carlo simulations are
	subsequently refined by local energy minimisation. For this step, we
	employ a finer parameter grid with a spacing of approximately $0.005$
	in both $\Delta_2$ and $\delta_1$. This procedure enables a more
	accurate determination of the stability regions of the various
	magnetic phases and yields the final phase diagram shown in
	Fig.~\ref{fig:PhaseDiag}. The resulting magnetic orders include
helical, spiral, and double-$\mathbf{Q}$ phases. Representative spin
textures are shown in Fig.~\ref{fig:Spintexture.SSF}, where the $S^x$
and $S^y$ components are represented by arrows and the $S^z$
component is encoded by the color scale. The cycloidal spiral state
shown in Fig.~\ref{fig:Spintexture.SSF}(a) can, to leading order, be
described as follows:
\begin{equation*}
	S^x_i\!\!	= \! S  \cos{(\textbf{Q}\textbf{R}_i)} , \mbox{~}	S^y_i\!=\! S  \sin{(\textbf{Q}\textbf{R}_i)}, 	\mbox{~}
	S^z_i	= 0,
\end{equation*}
where the propagation vector $\mathbf{Q} = (2\pi/7,2\pi/7)$.

The helix shown in Fig.~\ref{fig:Spintexture.SSF}(b) can be described
as
\begin{equation*}
	S^{x'}_i\!\!	= \! S  \cos{(\textbf{Q}\textbf{R}_i)} , \mbox{~}	S^{z'}_i\!=\! S  \sin{(\textbf{Q}\textbf{R}_i)}, 	\mbox{~}
	S^{y'}_i	= 0.
\end{equation*}
We use a rotated coordinate system $x'y'z'$ in which the $y'$ axis lies along the
propagation vector $\mathbf{Q}=(2\pi/7,2\pi/7)$, and the $x'z'$ plane is perpendicular
to $y'$. The propagation vector is close to the one observed experimentally.

Fig.~\ref{fig:Spintexture.SSF}(c) shows the double-\textbf{Q} phase.
Its magnetic structure factor is given in Fig.
\ref{fig:Spintexture.SSF}(f). It consists of two pairs of peaks at
the propagation vectors $\mathbf{Q} =(2\pi/7,2\pi/7)$ and
$\mathbf{Q'} = (-2\pi/7,2\pi/7 )$ with different magnitudes. In a
leading approximation, the spin configuration can be described as a
superposition of a helix and modulation in ($\pi,\pi$) direction,
characterized by propagation vectors $\mathbf{Q}$ and $\mathbf{Q'}$.
\begin{equation}
	\begin{split}
		S^{ x' }_i & = A_1 \sin{(\textbf{Q}\textbf{R}_i)},  \\
		S^{ y' }_i & = A_2 \sin{(\mathbf{Q'}\textbf{R}_i)}, \\
		S^{ z' }_i & = A_1 \cos{(\textbf{Q}\textbf{R}_i)}.  \\
	\end{split}
\end{equation}
Here, the parameters $A_1$ and $A_2$ determine the relative
amplitudes of the spin components along each direction. We found that
$A_2 \approx A_1 / \sqrt{3}$, the corresponding amplitude-squared ratio of $3$ matches
the intensity ratio of the two dominant peaks in Fig.~\ref{fig:Spintexture.SSF}(f), confirming that the out-of-plane
$S^{y'}$ modulation is subdominant. The ratio corresponds to
Phase~I in~\cite{Andriushin2024}.
 
\section{Magnetic structure factor}
Now we calculate the magnetic structure factor
\begin{equation}
	|{\cal F}_\mathbf{q}|^2 = \frac{1}{N}\sum_{\alpha=x,y,z}\left|\sum_i S^\alpha_i e^{-i\mathbf{q} \mathbf{R}_i} \right|^2,
\end{equation}
where $\mathbf{R}_i$ is the coordinate of the $i$-th spin.  For a perfectly ordered state, the magnetic static structure factor exhibits sharp $\delta$-function Bragg peaks at the ordering wavevectors precisely as shown in Fig. \ref{fig:Spintexture.SSF}(d–f). Panels (d) and (e) present the magnetic static structure factor for single-$\mathbf{Q}$ structures, while panel (f) shows the double-$\mathbf{Q}$ structure. In the latter case, the structure is characterized by two dominant peaks corresponding to the leading modulations discussed in the previous section, along with weaker sub-leading peaks of negligible intensity.

A single-$\mathbf{Q}$ helix is characterized by a single pair of Bragg peaks at $\pm\mathbf{Q}$. On the square lattice, however, the $C_4$ symmetry renders the two diagonal propagation vectors $\mathbf{Q}=(Q,Q)$ and $\mathbf{Q}'=(-Q,Q)$ equivalent. Consequently, a macroscopic sample generically fragments into domains in which the helix propagates along either diagonal. Since elastic neutron scattering adds these domains incoherently, the static structure factor of such a multi-domain single-$\mathbf{Q}$ state is the superposition of individual single-domain patterns and exhibits peaks at both $\mathbf{Q}$ and $\mathbf{Q}'$ at identical positions to those of the genuine double-$\mathbf Q$ order in Fig. \ref{fig:Spintexture.SSF}(f). Moreover, the relative peak intensities are determined by the (uncontrolled) domain populations, rendering the intensity imbalance an unreliable fingerprint. The static structure factor alone thus cannot distinguish a single-domain double-$\mathbf{Q}$ state from a multi-domain ensemble of single-$\mathbf{Q}$ domains. This ambiguity motivates the dynamical response analysis in the following sections.

\section{Magnon spectrum}

To find the excitations, the Holstein--Primakoff transformation is
applied~\cite{HP, Mattis}. For this purpose we introduce a local
coordinate system, such that the quantisation axis is parallel to the classical spin $\bm{S}_i$. Correspondingly, the spin operator transforms
$\hat{\mathbf{S}} \to\hat{\tilde{\mathbf{S}} }$. For notational
simplicity, we omit the operator symbol.

\begin{figure*}[!t]
	\begin{center}
		\includegraphics{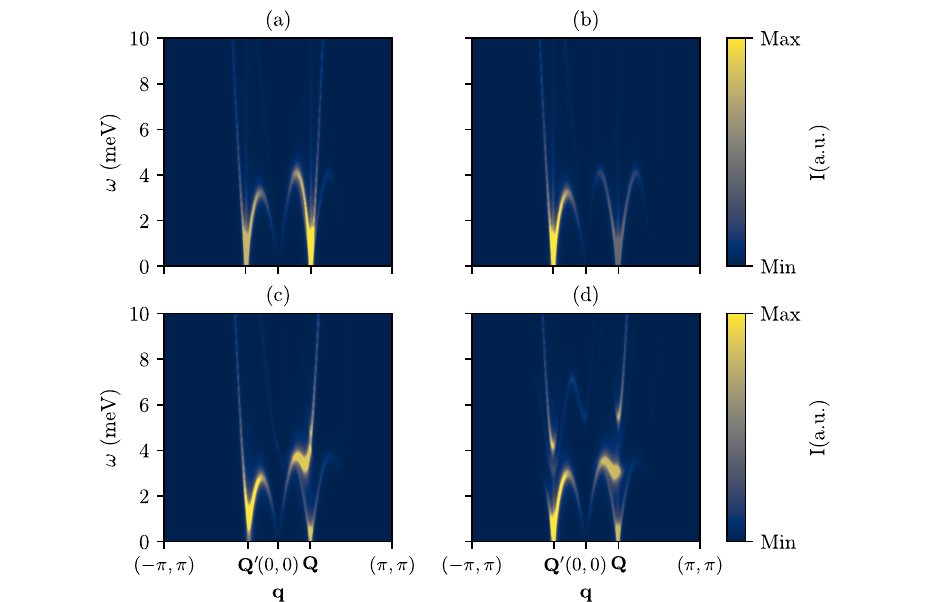}
	\end{center}
	\caption{
		Contour maps of the calculated inelastic neutron scattering cross-section $I(\mathbf{q},\omega)$ for the following magnetic structures:
		(a) Helix,
		(b) Spiral,
		(c) Helix with anisotropy,
		(d) Double-$\mathbf{Q}$. The parameters are the same as in Fig.~\ref{fig:Spintexture.SSF}.
			For the helix with anisotropy, we use the same exchange parameters as for the double-$\mathbf{Q}$ state, while the anisotropy parameters are $\delta_1 = -0.16$~meV and $\Delta_2 = -0.16$~meV.}\label{fig:dynamical.structure.factor}
\end{figure*}

Then we change the enumeration of the spins, introducing an internal
index $m$, labeling the position of the spin inside of the unit cell
and an external index $n$, which runs over the unit cells. Then the
Holstein--Primakoff yields
\begin{equation}
	\begin{split}
		\tilde{S}^x_{nm} & \to \frac{\sqrt{2S}}{2}(\hat{b}_{nm} + \hat{b}_{nm}^{\dagger})  \\
		\tilde{S}^y_{nm} & \to \frac{\sqrt{2S}}{2i}(\hat{b}_{nm} - \hat{b}_{nm}^{\dagger}) \\
		\tilde{S}^z_{nm} & \to S - \hat{b}_{nm}^{\dagger}\hat{b}_{nm}.
	\end{split}
	\label{eq:HP mapping}
\end{equation}

Noticing the translation invariance of the Hamiltonian we apply the
Bloch theorem, followed by the Bogoliubov transformation:
$\hat b_{\mathbf{q} m} = \sum_{ m' } U_{\mathbf{q}, mm' } \hat a_{\mathbf{q}
			m'} + V_{\mathbf{q}, mm' } \hat a^\dagger_{-\mathbf{q} m'} $ \cite{suppl}.
The matrices $ U_{\mathbf{q}, mm' }$ and $V_{\mathbf{q}, mm' }$ are
chosen so, that the new operators $a_{\mathbf{q} m'}$ satisfy
commutation relations for Bose particles.

Keeping only quadratic terms we get the quadratic in operators
Hamiltonian:
\begin{equation*}
	H = \sum_\mathbf{q} \sum_m \varepsilon_{\mathbf{q}m} \hat a_{{\mathbf{q}}m}^\dagger \hat a_{\mathbf{q}m}.
\end{equation*}

\section{Dynamical structure factor}

Now, we address the central question of our work: the dynamical response of the magnetic structures. 
At $T=0$ it can be obtained via 
 the fluctuation--dissipation relation, $$S_{\alpha\beta}(\mathbf q,\omega)
=\tfrac{1}{\pi}\,\chi''_{\alpha\beta}(\mathbf q,\omega)$$ 
for $\omega>0$. So only the imaginary part of the susceptibility $\chi''_{\alpha\beta}(\mathbf q,\omega)$ enters the inelastic neutron cross section, which takes the form \cite{Squires_2012}
\begin{equation}
	\begin{aligned}
		I(\mathbf{q},\omega) \sim - \sum_{\alpha \beta}\!\! \left(\delta_{\alpha \beta}
		- \frac{q_{\alpha}q_{\beta}}{q^2}\right) \chi''_{\alpha \beta}(\textbf{q},\omega).
	\end{aligned}
\end{equation}
The results for $I(\mathbf{q},\omega)$ in the unfolded Brillouin zone are presented in Fig. \ref{fig:dynamical.structure.factor}.

Figs. \ref{fig:dynamical.structure.factor}(a) and
\ref{fig:dynamical.structure.factor}(b) show the dynamical structure
factor for single-\textbf{Q} states, the helix and the cycloidal
spiral state. In both cases, the magnon dispersion
$\omega(\mathbf{q})$ vanishes at $\Gamma$-point and at the
propagation vector $\mathbf{Q}$. This behavior is consistent with the
Goldstone theorem~\cite{Goldstone1961}. In addition, it vanishes at
$\mathbf{Q}'$ due to the approximation of the linear spin waves of
the $1/S$ expansion, used in these calculations. However, it is known
that the gap at this point opens in higher orders of the $1/S$
expansion \cite{Chubukov1994,Tymoshenko2017}.

Figure~\ref{fig:dynamical.structure.factor}(c) shows the helix state
	in the presence of anisotropy. Compared with the isotropic helix in
	Fig.~\ref{fig:dynamical.structure.factor}(a), anisotropy lifts the
	degeneracy at $\mathbf{Q}$ and resolves the low-energy response into
	two branches. At $\mathbf{Q}'$, the overall structure remains similar
	to the isotropic case, including the pronounced outer vertical branch. But in contrast to the 
    isotropic case the anisotropy opens a finite gap of order $\sim 1~\mathrm{meV}$. The gap
	also reflects the absence of the second ordering vector, which in the
	double-$\mathbf{Q}$ state protects the gapless phason mode. The
	spectrum therefore retains the qualitative single-$\mathbf{Q}$
	character, but with anisotropy-induced mode splitting and gap
	opening.

Figure~\ref{fig:dynamical.structure.factor}(d) displays the
calculated inelastic neutron-scattering cross section of the
double-$\mathbf{Q}$ structure. At the $\Gamma$-point, one finds a
Goldstone mode whose spectral weight in the inelastic
neutron-scattering cross section vanishes, similarly to the helix
case. In contrast to the isotropic helix, however, an additional
roton-like mode appears at the $\mathbf{Q}$-point beside the
Goldstone mode. By analyzing the contributions to
$\chi_{\alpha\beta}$, we find that the phason mode, the mode
corresponding to in-plane phase shift of the magnetic helix,
preserves its Goldstone nature and remains gapless. At the same time
the spin-tilting mode, corresponding to tilting of the spins away
from the helix plane, becomes gapped. As well, a similar gapped
	mode appears in the anisotropic helix state. However, in that case
	the pronounced outer vertical branch remains clearly visible. At the
$\mathbf{Q}'$-point, the phason mode vanishes similar to helix.
However, the spectral weight of the outer branch is reduced in
comparison with the helix. The roton-like mode is shifted to higher
energies and becomes smeared out in comparison with the
$\mathbf{Q}$-point. Because Figs.~\ref{fig:dynamical.structure.factor}(c) and \ref{fig:dynamical.structure.factor}(d) use identical exchange constants and differ only in anisotropy
and in the magnetic structure, the comparison isolates the structural effect: the loss of the
outer vertical branch and the gapless phason are generic fingerprints of double-$\mathbf Q$ order,
whereas the gap magnitude and branch splitting are specific to the chosen parameters. Overall, the calculated cross section looks much
more similar to the experimental one for Sr$_3$Fe$_2$O$_7$ in the
relevant low-energy regime.

\section{Discussion and Conclusion}

In this paper, we study a double-$\mathbf{Q}$ structure on a square
lattice, which represents the simplest model allowing
single-$\mathbf{Q}$ and double-$\mathbf{Q}$ structures. The model
corresponds to Sr$_2$FeO$_4$, a system with a single magnetic layer
and a noncollinear magnetic structure \cite{Adler2022}. To date,
however, only elastic neutron-scattering experiments on
polycrystalline samples have been performed for this material.
At the same time, the most important outcome of our analysis, namely
 the prediction of a roton-like mode in a double-$\mathbf{Q}$ state, 
suppression of outer vertical branch at $\mathbf{Q}'$, 
goldstone mode at $\mathbf{Q'}$ is also relevant to the bilayer compound Sr$_3$Fe$_2$O$_7$. However,
because this system is bilayered, the $q_z$ dependence of the
dispersion becomes important. In particular, at the point $(0,0,\pi)$
the dispersion has a gap, in contrast to the point $(0,0,0)$. Indeed,
such a gap is observed experimentally \cite{Andriushin2024}, since
the measurements were performed at momentum transfer $(q,q,\pi)$
\footnote{Note that in the experiment in \cite{Andriushin2024} the
direction of the main propagation vector $\mathbf{Q}$ is along $[-1,1,\pi]$
in contrast to our single-layer convention $ \mathbf{Q}||[1,1,0]$. }.

Overall, the calculated double-$\mathbf Q$ spectrum is in qualitative agreement with several
characteristic features of the measured response of Sr$_3$Fe$_2$O$_7$ in the range from
$\Gamma$ to the $\mathbf{Q}'$ point (cf.\ Fig.~S1 of Ref.~\cite{Andriushin2024}). 
A fully quantitative
comparison would require a bilayer model including the $q_z$-dependent dispersion and the
experimental resolution function, which is beyond the scope of this work.
Since real crystals generically form multi-domain configurations, we have further verified that these features persist after domain averaging. As shown in Fig.~S2 of the Supplemental Material~\cite{suppl}, calculations for two-domain magnetic structures demonstrate that the characteristic signatures remain clearly resolved. 

In summary, we have investigated inelastic neutron scattering
cross-section for a set of magnetic configurations that can arise in
a two-dimensional square-lattice system. By constructing the phase
diagram as a function of the anisotropy parameters, we identified the
regions of parameter space in which the different magnetic structures
are stabilized. We then analyzed the corresponding dynamical
structure factors. We have demonstrated that the characteristic
feature of the double-$\mathbf{Q}$ structure in inelastic neutron
scattering is a roton-like magnon mode, suppression of outer vertical branch at $\textbf{Q}'$, goldstone mode at $\textbf{Q}'$ . Recent findings by M.
Mostovoy with collaborators
\footnote{M. Mostovoy, private
	communication.}
show similar results.

\section{acknowledgments}
We thank D. Inosov, D. Peets, M. Mostovoy, B. B\"uchner, G.
Khaliullin, O. Sushkov for fruitful discussions. We thank U.
Nietzsche for technical support. We acknowledge the financial support
of DFG (grant numbers 529677299, 455319354).

\bibliography{bibliography}
\newpage

\setcounter{equation}{0}
\setcounter{figure}{0}
\setcounter{table}{0}
\setcounter{page}{1}
\makeatletter
\renewcommand{\theequation}{S\arabic{equation}}
\renewcommand{\thefigure}{S\arabic{figure}}
\renewcommand{\bibnumfmt}[1]{[S#1]}
\renewcommand{\citenumfont}[1]{S#1}

\section*{Supplemental Material for:\\
	``Single-Q and Double-Q magnetic orders: A Theoretical Analysis of Inelastic Neutron Scattering in a Centrosymmetric Structure''}

\subsection{Monte Carlo Simulations}

To access complex multi-$\mathbf{Q}$ states in the parameter space of
$\delta_1$ and $\Delta_2$, where the Luttinger-Tisza method is not
applicable, we determine the classical ground state using Monte Carlo
simulations based on the Metropolis
algorithm~\cite{JankeMinimization}. This provides a numerical
framework for exploring noncollinear and multi-$\mathbf{Q}$ ordering
beyond analytical approaches.

We start with coarse Monte Carlo simulations on a discrete grid in
the $(\Delta_2,\delta_1)$ parameter space, using a step size of
$0.08$ for both parameters. For each parameter set, the simulation is
initialized from a random spin configuration and gradually cooled
from high to low temperatures. During the simulation, trial spin
updates are proposed and accepted or rejected according to the energy
difference between the current and proposed configurations. The
acceptance probability follows the standard Metropolis criterion:
\begin{equation} \label{eq: probability}
	P({\vec{x}_t \rightarrow \vec{x}_{t+1}}) =
	\begin{cases}
		1                              & \text{if } {E\left(\vec{x}_{t+1}\right) < E\left(\vec{x}_t\right)},    \\
		e^{\frac{-\Delta E}{T(t)}} & \text{if } \small{E\left(\vec{x}_{t+1}\right) \geq E\left(\vec{x}_t\right)},
	\end{cases}
\end{equation}

\begin{figure}[b]
	\begin{center}
		\includegraphics{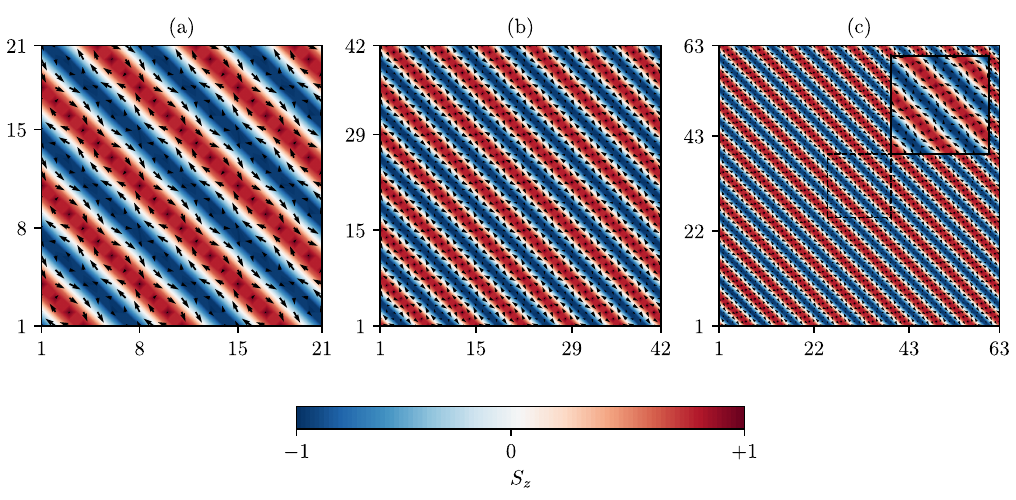}
	\end{center}
	\caption{Double-$\mathbf Q$ spin configuration obtained for the larger spin systems: $N=$ (a) 21 $\times$ 21, (b) 42 $\times$ 42, (c) 63 $\times$ 63 spins.}\label{fig: SpinTexture_21_42_63}
\end{figure}
where \( \Delta E = E(\vec{x}_{t+1}) - E(\vec{x}_t) \) represents the
energy difference between the current and proposed states, and \(
T(t) \) is the temperature at time step \( t \).

We gradually decrease the temperature according to the cooling
schedule:
\[
T(t) = T_0 \left( 1 - \frac{10}{N} \right)^t,
\]
where \( N \) is the total number of iterations, and \( T_0 \) is the
initial temperature. The temperature decreases such that its final
value is approximately $10^{-5}$.

The acceptance condition is given by:
\begin{equation}
	\text{if } \Delta E < 0 \text{ or } e^{-\Delta E / T} \geq \text{rand}(0,1),
	\label{MCacceptance_condition}
\end{equation}
where
\( \text{rand}(0,1) \)
is a random number uniformly distributed between 0 and 1.
After the annealing process, we perform
the L-BFGS optimization~\cite{NumericalOptimization} of our system in order
to end up in a local minima.

To demonstrate the stability of our solution and to rule out
finite-size effects, we repeated the calculations for larger lattices
(see Fig.~\ref{fig: SpinTexture_21_42_63}).

\subsection{Holstein-Primakoff Transformation}

In order to proceed with the Holstein-Primakoff transformation, we
choose a local coordinate system  with the quantization axis along
the classical spin. The spin component in the laboratory coordinate
systems and in the local systems are related as follows:
\begin{align}
	\begin{bmatrix}
		S^x_{{nm}} \\
		S^y_{{nm}} \\
		S^z_{{nm}} \\
	\end{bmatrix}
	= \begin{bmatrix}
		\cos \phi_{m} & -\sin \phi_{m} & 0 \\
		\sin \phi_{m} & \cos \phi_{m}  & 0 \\
		0             & 0              & 1 \\
	\end{bmatrix}
	\begin{bmatrix}
		\cos \theta_{m}  & 0 & \sin \theta_{m} \\
		0                & 1 & 0               \\
		-\sin \theta_{m} & 0 & \cos \theta_{m} \\
	\end{bmatrix}
	\begin{bmatrix}
		\tilde{S}^x_{{nm}} \\
		\tilde{S}^y_{{nm}} \\
		\tilde{S}^z_{{nm}} \\
	\end{bmatrix}.
\end{align}

Then after expansion and combining terms the Hamiltonian Eq. 1 of the main text takes the form:
\begin{equation}
	\begin{aligned}
		H = \sum_{\ell=1..3} & \sum_{nm, n'm'}^N  J_{\ell}^{xx}\tilde{S}_{nm}^x \tilde{S}_{n'm'}^x R^{xx}_\ell(\theta_{m},\phi_{m},\theta_{m'},\phi_{m'}) + J_{\ell}^{yy}\tilde{S}_{nm}^y \tilde{S}_{n'm'}^y R^{yy}_\ell(\theta_{m},\phi_{m},\theta_{m'},\phi_{m'}) + \\
		& + J_{\ell}^{zz} \tilde{S}_{nm}^z \tilde{S}_{n'm'}^z R_\ell^{zz}(\theta_{m},\phi_{m},\theta_{m'},\phi_{m'}) - J_{\ell}^{xy}\tilde{S}_{nm}^x \tilde{S}_{n'm'}^y R_\ell^{xy}(\theta_{m},\phi_{m},\theta_{m'},\phi_{m'}) -                 \\
		& -J_{\ell}^{xy}\tilde{S}_{nm}^y \tilde{S}_{n'm'}^x R_\ell^{yx}(\theta_{m},\phi_{m},\theta_{m'},\phi_{m'}) .
	\end{aligned}
	\label{eq: LocalHamiltonian}
\end{equation}
Note that for the next-nearest-neighbor ($J_2$) bonds, which lie along the diagonal,
the anisotropic exchange acquires off-diagonal ($xy$) components; the corresponding
cross terms $J^{xy}_{\ell}$ therefore appear in Eq.~\ref{eq: LocalHamiltonian}.

After this we can perform Holstein-Primakoff transformation in the
local coordinate system:

\begin{equation}
	\begin{aligned}
		\tilde{S}^x_{nm} & \to \frac{\sqrt{2S}}{2}(\hat{b}_{nm} + \hat{b}_{nm}^{\dagger}) ,  \\
		\tilde{S}^y_{nm} & \to \frac{\sqrt{2S}}{2i}(\hat{b}_{nm} - \hat{b}_{nm}^{\dagger}) , \\
		\tilde{S}^z_{nm} & \to S - \hat{b}_{nm}^{\dagger}\hat{b}_{nm}.
	\end{aligned}
\end{equation}

\subsection{Bogoliubov Transformation}
Consider a system of $N$ spins in a unit cell, where the Hamiltonian
has been expressed after applying the Holstein-Primakoff
transformation. The Hamiltonian takes the form:
\begin{equation}
	\begin{aligned}
		H = \sum_{\mathbf{q},mm'} & \, A^{mm'}_\mathbf{q} \hat{b}^\dagger_{\mathbf{q},m} \hat{b}_{\mathbf{q},m'} + A^{mm'}_{-\mathbf{q}} \hat{b}^\dagger_{-\mathbf{q},m} \hat{b}_{-\mathbf{q},m'}     \\
		& + B^{mm'}_\mathbf{q} \hat{b}_{-\mathbf{q},m'} \hat{b}_{\mathbf{q},m} + ( B^{m m'}_{\mathbf{q}})^* \hat{b}^\dagger_{\mathbf{q},m} \hat{b}^\dagger_{-\mathbf{q},m'}.
	\end{aligned}
\end{equation}
Note that the coefficients $A^{mm'}_\mathbf{q}$ and $B^{mm'}_\mathbf{q}$ satisfy the
relations: $A^{m'm}_\mathbf{q} = (A^{mm'}_\mathbf{q})^*$ and $B^{m'm}_\mathbf{q} =
B^{mm'}_{-\mathbf{q}}$.

To diagonalize the Hamiltonian, we use the Bogoliubov transformation,
which expresses the operators in terms of new quasiparticle
operators:
\begin{equation}
	\begin{aligned}
		a_{\mathbf{q}, m}          & = \sum_{m'=1}^{N} u_{\mathbf{q}}^{m,m'} \hat{b}_{\mathbf{q},m'} + v_{-\mathbf{q}}^{m,m'} \hat{b}^\dagger_{-\mathbf{q},m'},         \\
		a_{\mathbf{q}, m}^\dagger & = \sum_{m'=1}^{N} (v_{-\mathbf{q}}^{m,m'})^* \hat{b}_{-\mathbf{q},m'} + (u^{m,m'}_{\mathbf{q}})^* \hat{b}^\dagger_{\mathbf{q},m'}.
	\end{aligned}
	\label{eq: Bogoliubov transformation}
\end{equation}

The coefficients
$u^{n,m}_{\mathbf{q}}$ and $v^{n,m}_{\mathbf{q}}$ must satisfy the
following normalization condition, ensuring the consistency of the
transformation:
\begin{equation}
	\sum_{m'=1}^N \left[ u_{\mathbf{q}}^{m_1,m'} (u_{\mathbf{q}}^{m_2,m'})^* - v_{-\mathbf{q}}^{m_1,m'} (v_{-\mathbf{q}}^{m_2,m'})^* \right] = \delta_{m_1,m_2}.
\end{equation}

Next, diagonalization of the Hamiltonian is achieved by enforcing the
commutation relation:
\begin{equation}
	[H, a_{\mathbf{q},m}] = -\omega_{\mathbf{q},m} a_{\mathbf{q},m},
\end{equation}
where $\omega_{\mathbf{q},m}$ are the quasiparticle energy eigenvalues.

\subsection{Unfolding Procedure}
The unfolding is done through identifying of the Wannier functions in
the folded and unfolded Brillouin zones. We have the following
relation between the Bloch function operators and the Wannier
function for the unfolded case $$
\hat{b}_\mathbf{q}=\frac{1}{\sqrt{N}}\sum_{ i=1 }^N e^{i \mathbf{q
	}\mathbf{R}_i} \hat{b}_i $$ and for the folded case as $$
\hat{b}_{\mathbf{q} m} = \frac{1}{\sqrt{N_1}} \sum_{n=1}^{N_1}
e^{i\mathbf{q}( \mathbf{R}_n+\mathbf{r}_{m} )} \hat{b}_{nm}. $$ Here
we enumerate the operators with two indices external $n$ and internal
$m$, $N$ is the number of the sites and $N_1$ is the number of the
unit cells.

Then
\begin{equation}
	\begin{aligned}
		\hat{b}_{\mathbf{q}m} & = \frac{1}{\sqrt{N_1}}\sum_{n=1}^{N_1}e^{i\mathbf{q}(\mathbf{R}_n+\mathbf{r}_m)}\hat{b}_{nm} = \frac{1}{\sqrt{N_1}}\sum_{n=1}^{N_1}e^{i\mathbf{q}(\mathbf{R
			}_n+\mathbf{r}_m)}\hat{b}_{i}  = \\
		& = \frac{1}{\sqrt{N_1}}\sum_{n=1}^{N_1}e^{i\mathbf{q}(\mathbf{R}_n+\mathbf{r}_m)}
		\frac{1}{\sqrt{N}}\sum_{\mathbf{q'}}e^{-i\mathbf{q'}(\mathbf{R}_n+\mathbf{r}_m)}\hat{b}_{\mathbf{q'}}  = \frac{1}{\sqrt{M}}\sum_{\textbf{ G }}e^{-i \textbf{G} \mathbf{r}_m}\hat{b}_{\mathbf{q}+\textbf{G}},
	\end{aligned}
\end{equation}
where $\mathbf{G}$ are the reciprocal lattice vectors and $M = N/N_1$ is the number of sites per magnetic unit cell.

\subsection{Effect of Magnon Damping and Multi-Domain Effects}
\begin{figure}[H]
	\begin{center}
		\includegraphics{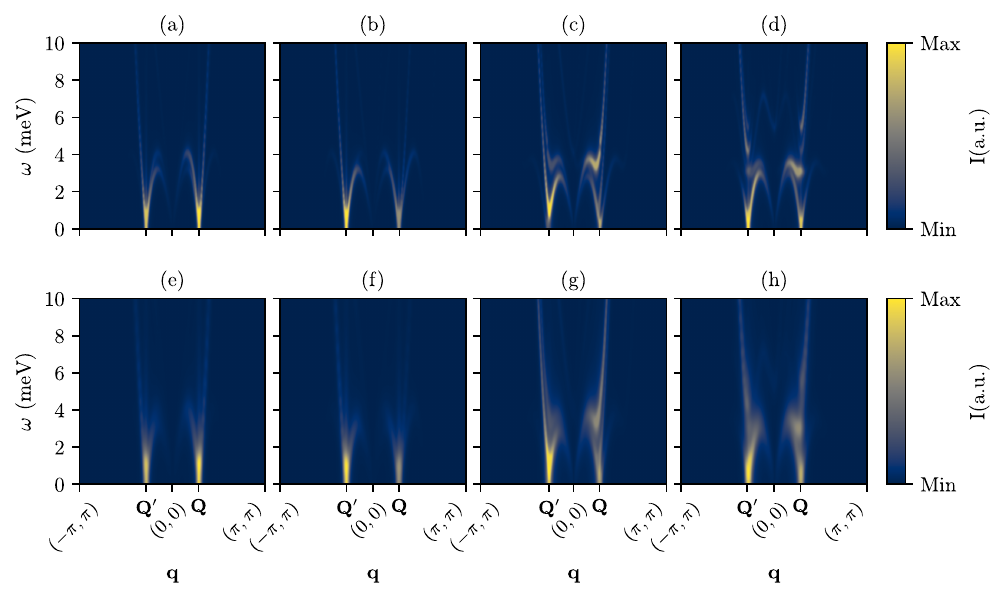}
	\end{center}
	\caption{Neutron scattering cross section including magnon damping and multidomain effects:
		(a) helix,
		(b) spiral,
		(c) helix with anisotropy,
		(d) double-\textbf{Q}, and
		(e-h) the corresponding two-domain spectra,
		obtained by combining the $C_4$-related domains in the population ratio $3:1$ and including a
		magnon damping $\gamma=0.6$~meV. Exchange parameters as in the main text.}\label{fig:DSFsmeared}
\end{figure}

In magnetic structures, magnon damping arises from several sources,
such as magnon–magnon interactions and the scattering of magnons on
impurities. To investigate the effect of magnon damping on the
neutron scattering cross section, we introduce a magnon-damping
parameter $\gamma$. Using $\gamma = 0.6$ meV, we calculated the
dynamical structure factor and the neutron scattering cross section.
To demonstrate the possible effect of multiple domains, we combine
the intensities in the ratio $3\!:\!1$. The result is shown in
Fig.~\ref{fig:DSFsmeared}. Comparing this result with the inelastic
neutron cross section without damping, we see that the roton mode is
smeared out but still clearly detectable. The spectrum along the
$(-\pi,\pi)$ direction becomes very similar to the experimental data
for Sr$_3$Fe$_2$O$_7$~\cite{Andriushin2024}.

\end{document}